\documentstyle[psfig]{l-aa} % 1993 style file for LaTeX
\def\cu{photon~cm$^{-2}$~s$^{-1}$~\AA$^{-1}$~sr$^{-1}$}

\def\ve{erg~cm$^{-3}$~s$^{-1}$~Hz$^{-1}$}
\def\cuhz{erg~cm$^{-2}$~s$^{-1}$~Hz$^{-1}$~sr$^{-1}$}

\begin{document} 

\thesaurus{11(11.05.2;  %Galaxies:evolution 
            11.09.3;  %Galaxies:intergalactic medium 
            11.19.5;  %Galaxies:stellar content
            11.17.1)}  %quasars:absorption lines
\title{
 The Lyman continuum radiation escaping from galaxies
}

%\subtitle{} 
\author { J.~-M.~Deharveng   \inst{}
          \and 
          S. Fa\"{\i}sse  \inst{}
        \and
          B. Milliard  \inst{}
           \and
             V. Le Brun \inst{}}
\offprints{J.~-M.~Deharveng, jmd@astrsp-mrs.fr}

\institute{Laboratoire d'Astronomie Spatiale du CNRS, Traverse~du~Siphon,
BP 8, 13376 Marseille Cedex 12, France.}

\date{Received date; accepted date} 

\maketitle 
\markboth{galaxy ionizing background}{}
\begin{abstract}
      The H$\alpha$ luminosity density of galaxies 
 is used to calculate
 the local diffuse radiation field at the Lyman limit 
   as resulting from star formation in the universe. 
   The fraction of Lyman continuum radiation leaking out from galaxies through
their neutral hydrogen is the most uncertain parameter of the calculation.
    A comparison of the diffuse radiation predicted from galaxies with
  that measured from all sources of
ionization shows that the average 
 Lyman continuum escape fraction should be lower than
 1\%.  This number is lower than the upper limits 
 reported so far in a few objects.
  It is also shown that the luminosity density at the Lyman limit resulting
  from star formation is consistent with the luminosity density and the 
  diffuse background measured in the far (non-ionizing) ultraviolet.

\end{abstract}

%_______________________________________________________________

\section{Introduction}
   
      It is thought that massive star formation in the universe may 
  provide a
  significant fraction of the background radiation that maintains
  the diffuse intergalactic medium and the 
  Lyman $\alpha$ forest clouds highly ionized (e.g. Bechtold et al. 1987,
  Songaila et al. 1990, Miralda-Escud\'e \& Ostriker 1990). This 
  contribution would augment that of quasars, especially at high redshift
  where their number density is observed to decline. The contribution of
  quasars itself depends on how our picture of their evolution is distorted
  by dust obscuration (Fall \& Pei 1993).

  Direct observations of the
  Lyman continuum (Ly$_{c}$) radiation escaping from galaxies 
  is however extremely difficult and, so far, only upper limits 
  have been obtained with the Hopkins Ultraviolet Telescope (HUT) 
  in four nearby star-forming galaxies (Leitherer et al. 1995).  
  Attempts to understand how Ly$_{c}$ radiation leaks out from 
  sites of star formation, 
  ionizes the diffuse interstellar medium around 
  and eventually escapes from a galaxy (e.g. Dove \& Shull 1994, 
  Patel \& Wilson 1995a,b, Ferguson et al. 1996)
  have shown that
 the phenomenon is dominated 
  by patchiness in the distribution
  of the neutral gas and should be highly random. Any quantitative 
  assessment of the contribution of galaxies to the ionizing background
  would therefore require a large number of observations before a 
  Ly$_{c}$ luminosity function is established. This uncertainty on the
  Ly$_{c}$ escape fraction is also a severe limitation for model
 predictions even though some of them are reasonably successful in linking
 the Ly$_{c}$ radiation produced by star formation to the rate of 
 chemical enrichment
 in the universe (Cowie 1988, Songaila et al. 1990, Madau \& Shull 1996).

  In this paper we use the recent determination of the H$\alpha$ luminosity 
 density of nearby galaxies by Gallego et al. (1995) to estimate the
 contribution of galaxies to the diffuse radiation background at the 
  Lyman limit and at $z=0$. 
 As a support to our approach, the H$\alpha$ luminosity density will be 
 compared  
 with other tracers of the 
  local star formation activity such as the luminosity density 
  and the diffuse background in the 
  far non-ionizing ultraviolet.  
  From the
 comparison between the 
  diffuse radiation at the Lyman limit predicted from 
  galaxies and that measured from all sources of ionization, we will derive 
   an upper limit   
   to the effective Ly$_{c}$ escape fraction in the local universe.

\section{Basic formulation}

  Following current formulation (e.g. Bechtold et al. 1987, 
  Meiksin \& Madau 1993)
  the mean specific intensity $J_{\nu_0}$ of the 
 diffuse radiation field at frequency $\nu_0$, as seen by an observer at
  redshift 0 (outside our Galaxy), writes as
 \begin{eqnarray} 
J_{\nu_0}&=& {c\over4 \pi H_0} \times\nonumber \\
 &&\int^{\infty}_{0} {1 \over (1+z)^3} 
 \epsilon(\nu,z)  {\exp\left[-\tau_{eff}(\nu_0,z)\right] \over (1+z)^2 (1+2q_0z)^{1/2}} 
\mathrm{d}z 
\end{eqnarray}
 where $\epsilon(\nu,z)$ is the proper volume emissivity (expressed in \ve)
  at frequency $\nu = \nu_0 (1+z)$ and redshift $z$, 
  and $exp[-\tau_{eff}(\nu_0,z)]$ is the mean transmission of a clumpy
  medium averaged over all lines of sight. 
  In our application the 
 volume emissivity is the ultraviolet emission resulting from 
the star formation activity in galaxies. Taking an average over all 
  morphological types and all star formation histories, the luminosity 
 density can be considered 
 continuous with a unique mean spectral shape. In these conditions 
 and following the notations of Bechtold et al. (1987),
 $\epsilon(\nu,z)$ can be split into 
  $\epsilon(\nu_0)$ the current local luminosity density of galaxies 
 at frequency
   $\nu_0$ and $k(\nu/\nu_0)$ the spectral shape normalized
  to 1 at the frequency $\nu_0$ with a factor                  
 $\psi(z)$ accounting for any proper
  evolution in the luminosity density (the density variation due to expansion
  is taken into account by a factor $(1+z)^3$).
  Adopting $q_0=0.5$, we get 
 \begin{eqnarray}
 J_{\nu_0}& =&  {c\over4 \pi H_0} \epsilon(\nu_0)\times\nonumber\\
&& \int^{\infty}_{0} 
{\psi(z)} \, k({\nu\over\nu_0}) 
 \, {1 \over(1+z)^{5/2}} \> {\exp\left[-\tau_{eff}(\nu_0,z)\right]} \mathrm{d}z
 \end{eqnarray}
   The effective optical depth $\tau_{eff}(\nu_0,z)$ due to 
 Lyman continuum absorption of H{\sc i} and He{\sc ii} by discrete absorption
 systems is given by (e.g. Paresce et al. 1980, M{\o}ller \& Jakobsen 1990, 
 Miralda-Escud\'e \& Ostriker 1990) 
 \begin{equation}
 \tau_{eff}(\nu_0,z)=\int^{z}_{0} \mathrm{d}z^\prime \int^{\infty}_{0} f(N,z^\prime)
 \,(1-\exp(-\tau))\,\mathrm{d}N
\end{equation} 
 where $f(N,z^\prime)$ is the redshift and column density distribution of
 absorbers along the line of sight
 and $\tau$ the optical depth through an individual cloud of column density
 $N$.

\section{Stellar contribution to the 
 radiation field at the hydrogen Lyman edge}

   We now use equation (2) to calculate the radiation background 
  at the Lyman limit $J_{912}$. The luminosity 
  density  is assumed to have a proper evolution parameterized 
  as $\psi(z)=(1+z)^{\gamma}$. Given the severe uncertainties about the spectral
  energy distribution of Ly$_{c}$ photons from star formation, 
  the normalized        
  spectral shape is modelled 
  as $k(\nu/\nu_0)=(\nu/\nu_0)^{\alpha}=(1+z)^{\alpha}$ 
  while we add a factor 
  for the absorption of stellar Ly$_{c}$ photons
  by the neutral hydrogen of each galaxy. 
  In order to account for the  $\nu^{-3}$
   frequency dependence of the absorption of Ly$_{c}$ photons by neutral
hydrogen,
 we have written the absorption factor as
   $exp[-6.3\times10^{-18} N_{H} (1+z)^{-3}]$. 
  The equivalent H{\sc i}
  column density term ($N_{H}$) is introduced for calculation purpose 
  (we make no assumption as to the geometry
and distribution of the gas) and the resulting effective Ly$_{c}$ escape 
  fraction
   will be defined as $f=exp[-6.3\times10^{-18}
N_{H}]$.
    In these conditions the background radiation J$_{912}$ (in \cuhz)
  at the Lyman limit 
   writes as
 \begin{eqnarray}
 J_{912}&=& {c\over4 \pi H_0} \epsilon(912) 
\int^{\infty}_{0} (1+z)^{\gamma+\alpha-5/2}\times\nonumber\\ 
&&\mathrm{e}^{-6.3\times10^{-18} N_{\mathrm{H}} (1+z)^{-3}}\,
\mathrm{e}^{-\tau_{\mathrm{eff}}(912,z)}\, \mathrm{d}z
\end{eqnarray}

\subsection {Adopted parameters}

 By comparison with existing models for the Ly$_{c}$ spectral energy
  distribution of star-forming population
   (e.g. Bruzual \& Charlot 1993), a value of $\alpha$ in the 
  range $-$1 to $-$3 is a realistic approximation. An upper bound to $z=2$ 
  is also realistic since it cuts any stellar flux contribution below 304~\AA~.
 The local luminosity density of galaxies $\epsilon(912)$ 
 at the Lyman edge 
   can be derived from
  the total H$\alpha$ luminosity
  per unit volume 
 of 1.26$\times$10$^{39}$ ergs s$^{-1}$ Mpc$^{-3}$
 evaluated for star-forming galaxies in the local universe 
  by Gallego et al. (1995).  Under the 
 current conditions valid in the ionized gas of galaxies 
  (T=10$^{4}$K and case B) we get
 a local density of 
  Ly$_{c}$ photons of 9.2$\times$10$^{50}$ s$^{-1}$ Mpc$^{-3}$ 
   (Osterbrock 1989). This number can be considered as a lower limit since 
  Ly$_{c}$ photons in optically thin gas produce fewer H$\alpha$ photons
  than those in optically thick gas.
 The relation between the 
 Ly$_{c}$ photons density and the luminosity density at 912~\AA~ depends on 
 the value of $\alpha$. For our average case $\alpha=-2$, we find 
 a luminosity density at 912~\AA\ of 
$4.8\times10^{37}$~erg~s$^{-1}$~\AA$^{-1}$~Mpc$^{-3}$  or
 $1.3\times10^{25}$~erg~s$^{-1}$~Hz$^{-1}$~Mpc$^{-3}$.  
  Incidentally, 
  the relation 
    $\log N_{\lambda}/L_{\lambda} = 13.28 \pm 0.16$ (photons~\AA\ erg$^{-1}$)
   established by Leitherer et al. (1995)  would give the same value.
 This relation 
  was established for starbursts with different star formation histories 
  and initial mass
 functions while the simplifying assumption of a continuous star formation rate
  is probably valid at the scale of the local universe. 

  Significant evolution of galaxies is now well established 
(e.g. 
  Ellis et al. 1996, Lilly et al. 1996, Fall et al. 1996) and we adopt
$\gamma=4$ from $z=0$ to $z=1$ as found 
  by Lilly et al. (1996) for the 
  evolution of the luminosity density of the universe 
   at 2800~\AA\. Insofar as 
  the light at this latter UV wavelength is essentially tracing 
  on-going star formation, we think that the same exponent should 
  be valid at our shorter wavelengths. Beyond $z=1$ the evolution is 
  known to slow down but the situation is less certain. We have adopted 
 $\gamma=0$, bearing in mind that 
 this choice is not critical since the contribution to the
 background at $z=0$ from objects at high redshifts is small 
 as soon as the evolution
is not strong. Last, the calculation is independent
  of the value of $H_{0}$ since luminosity densities scale as $H_{0}$.

\subsection {The intergalactic opacity term}

   At the Lyman edge and ignoring absorption due to He{\sc ii}
  for a line of sight limited to $z=2$ (He{\sc i} absorption is negligible),
  the effective optical depth $\tau_{eff}(912,z)$ in equation (4) writes as
\begin{eqnarray} 
\tau_{eff}(912,z)&=&\int^{z}_{0} \mathrm{d}z^\prime
 \int^{\infty}_{0} f(N,z^\prime)\times\nonumber\\
&&
  \left(1-\exp\left[-6.3\times10^{-18}(1+z^\prime)^{-3}N\right]\right)
  \,\mathrm{d} N  \nonumber
\end{eqnarray}
  Assuming a power-law of exponent $-1.5$ for the column density distribution 
 (Petitjean et al. 1993, Songaila et al. 1995), 
 and adopting the line densities per unit redshift and the 
 evolution parameters from Boksenberg (1995), $f(N,z^\prime)$ writes as
   $ 1.07\times10^{8} N^{-1.5} (1+z^\prime)^{0.58}$
   for the Lyman $\alpha$ forest clouds,
 and $5\times10^{7} N^{-1.5} (1+z^\prime)^{1.5}$ 
 for the Lyman limit systems ($N > 1.6\times10^{17}$ cm$^{-2}$). 
The calculation of the first  normalization constant
 accounts for the  detection limit of 0.24~\AA\ rest equivalent width and
 a velocity width of 30 km s$^{-1}$ as in Miralda-Escud\'e \& Ostriker (1990).
 The parameters for the Lyman $\alpha$ forest clouds have been obtained 
 for $0<z<1.3$ but the plot of their evolution up to $z=3.7$ (Boksenberg 1995)
 shows that our parameterization remains appropriate till the 
  adopted limit at $z=2$.
  As a numerical example, we find a 
transmission from $z=0$ to $z=2$
    $exp(-\tau_{eff}(912,2))=0.384$. 
 Playing with
the error bars given on the line densities per unit redshift and the
 evolution parameters 
 (Boksenberg 1995) we find that this transmission 
 does not change by more
 than 40\%.

\subsection {Results}

   The background radiation calculated by equation (4) 
   with the average transmission models discussed above 
   is displayed in Figure 1 
   as a function of the Ly$_{c}$ escape fraction.  Although our evaluation
  is based on the measured Ly$_{c}$ photons density in the local universe 
  and avoids
  therefore most of the uncertainties inherent to pure model calculations,
 it still depends on a few parameters,
 the index $\alpha$ of the average 
  spectral shape in the Lyman continuum, the evolution factor 
 and the opacity of the
  intergalactic medium.  The resulting uncertainties are illustrated 
 in Fig. 1. 
  First, the impact of the ill-known 
  index $\alpha$ (values $-1, -2, -3, -4$ are used in Fig. 1), is found to be
 reduced 
 by the relation between $\alpha$ and 
  the luminosity density at the Lyman edge for a given 
 Ly$_{c}$ photon density.
   Second, the effect of a larger intergalactic opacity 
  as obtained with the upper 
  limits given by Boksenberg (1995) on the density of Lyman clouds and 
  Lyman limit systems per unit redshift is modest and comparable 
  with a change of one unit of the index $\alpha$.
   In contrast, the calculation is sensitive to the amount of 
   evolution as shown with the case of a milder evolution $\gamma=2$ 
  (till $z=2$) plotted in Fig. 1.
  Selected as the variable against which the diffuse radiation has been 
  plotted in Fig. 1, the Ly$_{c}$ escape fraction is,
  as anticipated, the major source of 
  uncertainty. We note, however, that 
  the diffuse radiation does not decrease as fast as the 
 Ly$_{c}$ escape fraction. 
 The main reason is that galaxies at high redshift contribute to the 
 diffuse radiation and may still be optically thin at $912/(1+z)$~\AA\
  while their nearby counterparts are optically thick at 912~\AA.
  The issue of the 
  uncertainty on the luminosity density itself at the Lyman edge
  will be addressed in the two following sections.
  
\begin{figure}
\centerline{\psfig{figure=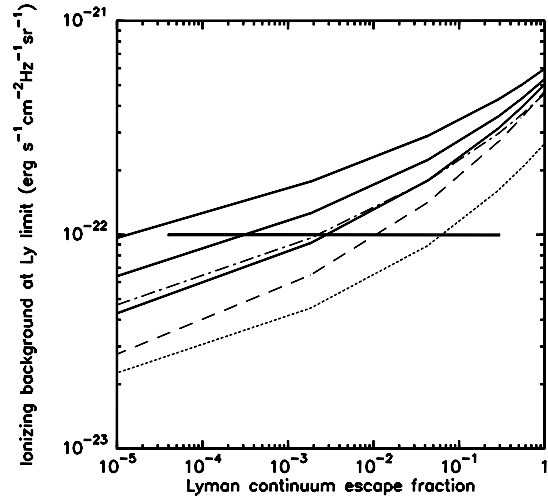,height=8cm,clip=t,angle=-90}}
\caption[]{Background radiation due to star formation in galaxies 
 predicted  at 
 the Lyman limit and $z=0$ as a function of the Ly$_{c}$ escape fraction.
 Solid lines: spectral index ($\nu^{\alpha}$, see text)  
 $\alpha = -1, -2, -3$ respectively from top to bottom.
 Long-dashed line: $\alpha = -4$.
 Dot-dashed line: $\alpha = -2$ and intergalactic opacity higher than 
 nominal with a density per unit redshift
 $(\mathrm{d}N/\mathrm{d}z)_{0}=30.9$ instead of 24.3 for the Ly forest clouds and 
 a density  per unit redshift
 $(\mathrm{d}N/\mathrm{d}z)_{0}=0.42$ instead of 0.25 for the Lyman limit systems.
 Short-dashed line: $\alpha= -2$ and $(1+z)^{2}$ evolution from $z=0$ to
 $z=2$ instead of 
  $(1+z)^{4}$ from $z=0$
 to $z=1$ and no evolution from $z=1$ to $z=2$ in all other cases. 
 The horizontal thick line is the upper limit to current measurements 
 of the diffuse background radiation at the Lyman limit and $z=0$.
}
\end{figure}

\section {Comparison with the ionizing background: constraints on the 
Ly$_{c}$ escape fraction}

   We have compared the predictions of Figure 1 
 with the measurements of the ionizing background 
  (at $z\approx 0$) at the Lyman edge.
 These measurements are only indirect and include the contribution of all
  ionizing sources, chief among them the quasars. They are therefore
  upper limits, possibly very generous, to the contribution of 
  galaxies. The measurements have been obtained so far through 
  a variety of methods (Bechtold 1993, for a review), 
  such as the H$\alpha$ emission from 21-cm 
  emitting clouds, the modeling of the sharp edges of H{\sc i} clouds
  and,  the proximity effect (Kulkarni \& Fall 1993),
 following the 
  observations of the Lyman $\alpha$ forest at low redshift with the HST.
  From the recent works of Dove \& Shull (1994), Vogel et al. (1995)
  and Donahue et al. (1995)    
  and review of previous determinations
  therein, 
  we conclude that the ionizing background at 900~\AA\
  should lie in the range  $0.06 - 1 \times10^{-22}$~\cuhz.
  At its lower bound the ionizing background is comparable to the 
  evaluations currently  made for the contribution of quasars
  (e.g. Madau 1992), indicating 
  a possible negligible role of the galaxies.

  Fig. 1 shows that, under most of the conditions,
   the  Ly$_{c}$ escape fraction is smaller than 1\%. This is
  smaller than the upper limits in the four star-forming galaxies observed
  by Leitherer et al. (1995), especially 
 after the modifications advocated by Hurwitz et al. (1997) on the basis of 
  unaccounted absorption by H{\sc i} gas in our Galaxy.
                    Our upper limit would be less restrictive than 1\% if
  the spectral index $\alpha$ is lower than $-4$, or the evolution milder 
  ($\gamma < 4$) than
 found by   Lilly et al. (1996). 
Insofar as our comparison
   was made with the intergalactic radiation field of all origin, our 
   upper limit is probably generous but, even if small, does not exclude 
  that galaxies make
   most 
   of the ionizing intergalactic radiation field.

  Our conclusion also depends crucially on 
 the value of the total H$\alpha$ luminosity density
  of the local universe in the sense that the larger the H$\alpha$ luminosity
density (or 
 the associated number of Ly$_{c}$ photons), the stronger is 
 the upper limit for the  Ly$_{c}$ escape
fraction. Our limit of 1\% seems firm 
 for two reasons. 
  First, the 
  H$\alpha$ luminosity function built by Gallego et al. (1995) is based on 
 star-forming galaxies with H${\alpha}+$N[II] equivalent widths 
 larger than 10~\AA\ 
  and therefore should more likely lead to an underestimation of the total
 H$\alpha$ luminosity density. Second, our conversion of the 
 H$\alpha$ luminosity density into a Ly$_{c}$ photon density under case B
 assumptions provides, as previously said, a lower limit on the latter quantity.
 
\section{Consistency with the non-ionizing far UV radiation}
 
     As the H$\alpha$ luminosity density of the local universe is a key
 input in our calculation, 
 we have tried to evaluate how this quantity and 
 the associated number of Ly$_{c}$ photons compare with 
 other measured quantities tracing the star formation activity in the universe
 such as the far-UV luminosity
 density as determined from surveys of galaxies
 or the far-UV (non ionizing) background as
 measured by in-orbit experiments.

  The former quantity is related to
the luminosity density emitted at 900~\AA\ by star formation activity 
 through two parameters, the Lyman break of a pure 
 star-forming population 
  and the extinction at far-UV wavelengths due to dust mixed with the young
 stars in each galaxy. The far-UV spectral energy distribution (from longward
of the Lyman break to $\approx$ 2000~\AA) of a pure (without dust) 
 and continuous star-forming population can 
 be assumed to be flat in energy per frequency unit as shown by the models
 of Bruzual \& Charlot (1993) and the observations of star-forming galaxies
 with little extinction by  Calzetti et al. (1994).
 In the case of a continuous rate as we expect
 for the average star formation
 in the local universe,
 the Lyman break factor should be 
 of the order of 4 according to Bruzual \& Charlot (1993) and possibly 
  between 6 and 20 depending on the initial mass function
  for massive stars according to
 Leitherer \& Heckman (1995). With an average far-UV extinction of the order
 of 1 mag as discussed by Deharveng et al. (1994), we get a
  far-UV (observed) to 900~\AA\ (emitted)  luminosity density ratio 
  between 1.6 and 8. For comparison, the 1600~\AA\ (observed) to 900~\AA\ 
 (emitted) flux ratios in the four galaxies observed by Leitherer et al. (1995) 
  are found to be 0.3, 1.5, 1.6 and 6. With the ratio above and our 
 local luminosity density $\epsilon(912)$ at the Lyman edge of
 $1.3\times10^{25}$~erg~s$^{-1}$~Hz$^{-1}$~Mpc$^{-3}$ we get
 a far-UV luminosity density  of  
  2 -- 10 $\times10^{25}$~erg~s$^{-1}$~Hz$^{-1}$~Mpc$^{-3}$. This is in 
agreement with the local
  luminosity density of 3$\times10^{18}$~W~Hz$^{-1}$~Mpc$^{-3}$ 
(H$_0$=50~km~s$^{-1}$~Mpc$^{-1}$)
  evaluated at 2800~\AA\ by Lilly et al. (1996) and of 
6$\times10^{18}$~W~Hz$^{-1}$~Mpc$^{-3}$ (H$_0$=50~km~s$^{-1}$~Mpc$^{-1}$) 
evaluated at 2000~\AA\ by Milliard et al. (1997). 
  
   Although the far-ultraviolet background is rich of several possible 
 components, some of them of galactic 
 origin, various arguments 
 have established the
 accumulation of galaxy light along the line of sight as the dominant
 extragalactic contributor (e.g. Bowyer 1991, Jakobsen 1995, for a review)
 with 
 an intensity in the  range 50 -- 150 \cu  
  (or 0.7 -- 2 $\times$10$^{-21}$ \cuhz\ at say 2000~\AA). 
   This background radiation can be converted back into a 
  local luminosity density $\epsilon(2000)$ 
  according to equation (2) which, 
  with the same assumptions as above but without the neutral gas 
  absorption term and the intergalactic opacity term, 
  writes as 
 \begin{equation}
 J_{2000}= {c\over4 \pi H_0} \epsilon(2000) \int^{1.2}_{0} 
 (1+z)^{\gamma+\alpha-5/2} \> \mathrm{d}z
 \end{equation}
 The integral upper bound is now 1.2, the value which shifts 
 the Lyman break at 2000~\AA. 
   Assuming $\gamma=4$ and $\alpha$ in the range $-$1 to $-$3 (the case 
  $\alpha=0$ is observed for galaxies with little extinction and is not 
  appropriate for the average spectral shape) the upper limit background of
  2$\times$10$^{-21}$ \cuhz\ gives a far-UV luminosity density in
  the range 2.6 -- 6 $\times10^{18}$
  W Hz$^{-1}$ Mpc$^{-3}$, again in satisfying agreement with the range of
  values that we have derived above.

\section{Conclusion}

   We have used the H$\alpha$ luminosity density of galaxies in 
 the local universe 
  for estimating the contribution 
  of galaxies to the ionizing background at $z=0$.
   This approach reduces the number of
  theoretical assumptions entering this type of evaluation
 and is therefore expected to narrow the range of predicted values. 
 Of the remaining
  parameters needed for the evaluation,
   the fraction of Ly$_{c}$ radiation leaking out from galaxies is 
   the most uncertain.
  From a comparison between the 
  ionizing photons predicted by on-going star formation in galaxies
  and those measured in the intergalactic radiation field,  it is found that 
  the Ly$_{c}$ escape fraction is smaller than 1\% in the 
  local universe. This average upper limit is more restrictive than
   the upper limits reported in four galaxies by Leitherer et al. (1995) 
   and re-analyzed by Hurwitz et al. (1997). 
   
%\begin{acknowledgements}
%  
%\end{acknowledgements}

% FIGURES
% Fig.1

\end{document}